%% file: CogSci_Template.tex
\documentclass[10pt,letterpaper]{article}

\usepackage{cogsci}

\cogscifinalcopy 

\usepackage{pslatex}
\usepackage{apacite}
\usepackage{float} 

\usepackage{amssymb}
\usepackage{amsmath}
\usepackage{booktabs}
\usepackage{multirow}
\usepackage{graphicx}
\usepackage[hyphens]{url}

\title{Improving Unsupervised Task-driven Models of Ventral \\Visual Stream via Relative Position Predictivity}

\author{
  {\large \bf Dazhong Rong$^{1}$\quad Hao Dong$^{2}$\quad Xing Gao$^{3}$\quad Jiyu Wei$^{1}$\quad Di Hong$^{1}$}\\
  {\large \bf Yaoyao Hao$^{1}$\quad Qinming He$^{1}$\quad Yueming Wang$^{1}$\thanks{Corresponding author.}} \\
  $^{1}$Zhejiang University, Hangzhou, China\\
  \texttt{\{rdz98,weijiyu,hongd,yaoyaoh,hqm,ymingwang\}@zju.edu.cn}\\
  $^{2}$Rutgers University, New Jersey, USA\\
  \texttt{hd322@scarletmail.rutgers.edu}\\
  $^{3}$University of Electronic Science and Technology of China, Chengdu, China\\
  \texttt{2022050903029@std.uestc.edu.cn}
}

\begin{document}

\maketitle

\begin{abstract}
Based on the concept that ventral visual stream (VVS) mainly functions for object recognition, current unsupervised task-driven methods model VVS by contrastive learning, and have achieved good brain similarity. However, we believe functions of VVS extend beyond just object recognition. In this paper, we introduce an additional function involving VVS, named relative position (RP) prediction. We first theoretically explain contrastive learning may be unable to yield the model capability of RP prediction. Motivated by this, we subsequently integrate RP learning with contrastive learning, and propose a new unsupervised task-driven method to model VVS, which is more inline with biological reality. We conduct extensive experiments, demonstrating that: (i) our method significantly improves downstream performance of object recognition while enhancing RP predictivity; (ii) RP predictivity generally improves the model brain similarity. Our results provide strong evidence for the involvement of VVS in location perception (especially RP prediction) from a computational perspective.

\textbf{Keywords:}
ventral visual stream; task-driven models; unsupervised learning
\end{abstract}

\input{sections/Introduction}
\input{sections/Background}
\input{sections/Methods}
\input{sections/Results}
\input{sections/Discussion}

\section*{Ethics Statement}
The biological datasets utilized in this study are either publicly accessible or obtained from published papers with the authors’ consent.
All data usage complies with respective data sharing and ethical guidelines.
No new data involving human or animal subjects were collected for this study.
\section*{Acknowledgments}
This work was supported in part by STI 2030—Major Projects (2022ZD0208903), in part by the National Natural Science Foundation of China under Grant 62336007, in part by the Starry Night Science Fund of the Zhejiang University Shanghai Institute for Advanced Study under Grant SN-ZJU-SIAS-002. Yueming Wang is the corresponding author.

\clearpage
\bibliographystyle{apacite}

\setlength{\bibleftmargin}{.125in}
\setlength{\bibindent}{-\bibleftmargin}

\bibliography{CogSci_Template}

\end{document}

%% file: sections/Introduction.tex
\section{Introduction}\label{sec:1}
Ventral visual stream (VVS) is a neural pathway in the brain primarily involved in the processing of visual information~\cite{cloutman2013interaction,zachariou2014ventral,simic2017dorsal}, and its disruptions are linked to various visual disorders (\textit{e.g.,} visual agnosia)~\cite{greene2005apraxia}.
Task-driven methods~\cite{nayebi2018task,nayebi2021goal,yamins2016using}, which aim to train artificial neural networks (ANNs) on a specific behaviorally relevant task rather than directly fitting neural data~\cite{wei2024speed}, are widely adopted to model VVS and have achieved considerable success.
The outputs of intermediate layers of ANNs trained by these task-driven methods demonstrate high similarity to the neural responses in VVS, where the higher similarity indicates the architecture of the ANNs or the functionality of the learning task is more brain-like.
Leveraging these task-driven methods, on the one hand, researchers can gain deeper insights into the potential neural mechanisms underlying the visual disorders; on the other hand, researchers can draw inspiration from biological vision to design ANNs that are more robust against adversarial examples~\cite{zhu2024multiview} and backdoor attacks~\cite{rong2024clean}.

Existing task-driven methods to model VVS can be roughly divided into two categories: supervised and unsupervised. Because of the better biological plausibility, recently unsupervised task-driven methods have received more considerable attention.
Among these unsupervised methods, contrastive learning~\cite{zhuang2021unsupervised} achieves the best brain similarity in modeling VVS.
However, we argue there is still room for improvement because the correspondence between the tasks of ANNs and the functions of VVS is not adequately considered.

In this paper, we conduct in-depth analysis of the correspondence in modeling VVS by contrastive learning.
We first consider the core function of VVS: object recognition.
Contrastive learning can endow the model with the capability of object recognition.
The evidence is that after training the model by contrastive learning, the image features extracted by the model are linearly separable for object category~\cite{chen2020simple,newell2020useful,ericsson2021well}.
However, object recognition is not the only function of VVS.
Some studies~\cite{konen2008two,kravitz2010high,sereno2011population,zachariou2014ventral} have suggested that the functional specialization of ventral and dorsal visual streams is not as distinct as originally proposed, with VVS also contributing to location perception, despite location perception is primarily performed by dorsal visual stream.
Hence, in this paper we distill location perception into relative position (RP) prediction and additionally consider this function in modeling VVS.
Unfortunately, because the constraints induced by contrastive learning are relative-position-agnostic, contrastive learning may be unable to endow the model with the capability of RP prediction.

To compensate for the above functional defect, in this paper we design RP learning which makes the model explicitly learn the RP between two sub-images, and propose a new unsupervised task-driven method to model VVS, combining contrastive learning and RP learning.
Our method can handle the functions of object recognition and RP prediction simultaneously.
The main contributions of this paper can be summarized as following:
\begin{enumerate}
    \item We analyze the correspondence between two unsupervised tasks (contrastive learning and RP learning) and two functions (object recognition and RP prediction) of VVS in depth. To the best of our knowledge, we are the first to consider the function of RP prediction in modeling VVS.
    \item We propose a new unsupervised task-driven method to model VVS, which combines contrastive learning and RP learning. Extensive experiments demonstrate that our proposed method sets the new state-of-the-art brain similarity of unsupervised task-driven models of VVS.
    \item Furthermore, our experimental results show RP predictivity not only enhances downstream performance of object recognition, but also improves the model brain similarity to various cortical regions (V1, V2, V4, IT) on VVS, proving VVS is indeed involved in the function of RP prediction.
\end{enumerate}

%% file: sections/Background.tex
\section{Related Work}
Some recent work~\cite{yamins2014performance,khaligh2014deep} has found that when achieving better performance of artificial neural networks (ANNs) on certain behaviorally relevant tasks (\textit{e.g.,} object recognition), the representations of ANNs' intermediate layers simultaneously become more similar to the recorded neural activities of primate VVS.
Based on this finding, various task-driven methods~\cite{gucclu2015deep,cichy2016comparison,kubilius2016deep,yamins2016using}, which optimize ANNs for better task performance rather than fitting the recorded neural activities directly, are proposed to model primate VVS and obtain great success.
The higher similarity between the intermediate outputs of ANNs and the neural activities of VVS indicates the model architecture or the task functionality is more brain-like.

Existing task-driven methods for modeling primate VVS can be categorized according to whether the task is supervised or unsupervised.
In the following, we summarize two sorts of work respectively.

\subsection{Supervised Task-Driven Methods}
VVS consists of a cascade of cortical areas which are hierarchically organized and anatomically distinguishable.
The low-level visual features (\textit{e.g.,} local orientation and spatial scale) are captured in the primary visual area V1, and then are processed into some more complex features in the mid-level visual areas V2, V3 and V4~\cite{carandini2005we,sincich2005circuitry,dicarlo2012does,schiller1995effect}.
Lastly, the most high-level features with linear separability of object category are obtained in the highest visual area IT (inferior temporal)~\cite{brincat2004underlying,yamane2008neural,hung2005fast}.

Based on the above theories about the emergence of the capability to recognize objects in the brain, existing supervised task-driven methods commonly adopt object recognition as the learning task and differ in other settings.
\cite{shi2019comparison} trains VGG-16, a multi-layer convolutional neural network (CNN), to model mouse visual cortex.
\cite{cadena2019deep} trains VGG-19 to model macaque V1 area.
For better matching the primate VVS, \cite{kubilius2019brain} specially designs a shallow recurrent CNN named CORnet-S consisting of four areas anatomically mapped to V1, V2, V4, and IT, and then trains CORnet-S to model macaque visual cortex.
Different from conventional ANNs, spiking neural networks (SNNs) encode information with time sequences of spikes, and hence act more like biological neurons.
For this better biological plausibility, \cite{huang2023deep} trains SEW ResNet, a type of SNNs, to model both mouse visual cortex and macaque visual cortex.

\subsection{Unsupervised Task-Driven Methods}
Despite the great success in achieving high brain similarity, all of the above supervised studies fail to explain: how the representations in the brain are learned?
These studies without exception assume the brain can receive both massive visual stimuli and the corresponding category labels, hence the supervised training.
However, factually there is a lack of convincing evidence to prove that the brain receives these huge numbers of category labels, especially for nonhuman primates and human infants.
This implausible assumption may cause inaccuracies in subsequent research on neural mechanisms of primate visual system.

Compared to supervised learning, unsupervised learning is more in line with the real situation.
In the brain, the capability to recognize objects is not obtained by training the neurons all the time to correctly categorize the input visual stimuli, but rather it emerges naturally through some unsupervised training tasks~\cite{zhuang2021unsupervised}.
This is consistent with recent findings in the field of computer vision, which suggest that pre-training on some unsupervised tasks can lead to better performance on downstream supervised tasks~\cite{chen2020simple,he2019rethinking,erhan2010does}.
There have been several studies devoted to modeling VVS through unsupervised task learning.
\cite{higgins2021unsupervised} trains a beta-variational auto-encoder ($\beta$-VAE) on image reconstruction task to model macaque IT area.
\cite{lotter2020neural} trains a recurrent generative network (PredNet) on the task of predicting future video frames to model macaque visual cortex.
\cite{zhuang2021unsupervised} trains ResNet-18 by SimCLR, a classic framework for contrastive learning, to model macaque V1, V4, and IT areas.
Note that it achieves the best brain similarity among the above studies.

However, unsupervised task-driven modeling of VVS is still under-explored.
Most of existing studies just try various unsupervised tasks to pursue higher similarity score between the intermediate outputs of ANNs and the neural activities recorded in visual cortex, but lacks consideration of the correspondence between the unsupervised tasks of ANNs and the functions of VVS.  

%% file: sections/Methods.tex
\section{Methods}\label{sec:method}
Modeling VVS involves two stages: \textbf{(i)} training a base model on image dataset, and \textbf{(ii)} evaluating the brain similarity of the trained model to four different cortical regions (V1, V2, V4, IT) of VVS on neural dataset.
In this section, we focus on the first stage and present our proposed unsupervised task-driven method to train the base model in detail.

\subsection{Preliminaries}
The base model takes images as inputs and extracts their features as outputs.
Formally, we denote the high-dimensional Euclidean space of input images by $\mathbb{M}^{c\times k\times k}$, where $c$ and $k$ respectively represent the number of channels and the number of pixels along each side of the images.
Besides, we denote the $n$-dimensional Euclidean space of the extracted feature vectors by $\mathbb{V}^n$.
Then the base model can be treated as a learnable function $f: \mathbb{M}^{c\times k\times k}\rightarrow\mathbb{V}^n$.

In each round of training, a batch of images denoted by $\{\mathbf{M}_i\}_{i=1}^{N}$ are randomly sampled from image dataset, where $\mathbf{M}_i$ is the $i$-th image in the current batch and $N$ is the batch size.
In order to model VVS by unsupervised learning, we need to design a specific unsupervised learning task, indicated by a loss function $\mathcal{L}$.
Then in each training round, we optimize the learnable base model $f$ by minimizing the loss $\mathcal{L}(\{\mathbf{M}_i\}_{i=1}^{N})$.

\subsection{Contrastive Learning}
Since object recognition is the core function of primate VVS~\cite{zhuang2021unsupervised,huang2023deep}, the first goal of our designed unsupervised learning task is to endow the base model with the capability to recognize objects.
Recent work~\cite{chen2020simple} proposes SimCLR, one of the most widely used frameworks for contrastive learning (CL), and demonstrates that after pre-training by SimCLR, a simple linear classifier trained on top of the frozen learned features can achieve significantly high image classification accuracy.
This indicates the features learned by SimCLR are linearly separable for object category, proving that the base model has obtained the capability to recognize objects.
This is also very consistent with the biological reality that the low-level visual features are progressively processed into the high-level visual features containing linear separability of object category through primate VVS~\cite{yamane2008neural}.
Due to the above facts, we adopt SimCLR to train our base model.

Specifically, in each round of training, $2N$ variants are generated from the $N$ images through data augmentations (\textit{e.g.,} crop, flip and color distortion), denoted by $\{\mathbf{X}_i\}_{i=1}^{2N}$.
Note that $\mathbf{X}_{2i-1}$ and $\mathbf{X}_{2i}$ are the variants of $\mathbf{M}_{i}$.
An additional learnable projection head denoted by $g:\mathbb{V}^n\rightarrow\mathbb{V}^m$ is involved in SimCLR, which is a multi-layer perceptron (MLP) and projects the features extracted by our base model into the $m$-dimensional vector space.
Based on our base model $f$ and the projection head $g$, we define the similarity between two variants as $\textbf{sim}(\mathbf{X}_i,\mathbf{X}_j)=\frac{g(f(\mathbf{X}_i))^{\text{T}}g(f(\mathbf{X}_j))}{\|g(f(\mathbf{X}_i))\| \|g(f(\mathbf{X}_j))\|}$, and define $\tilde{s}_{i,j}=\frac{\exp(\textbf{sim}(\mathbf{X}_i,\mathbf{X}_j))}{\tau}$, where $\tau$ is the temperature parameter.
Furthermore, we define $\tilde{s}_i=\sum_{j=1}^{2N}\tilde{s}_{i,j}-\tilde{s}_{i,i}$.
Finally, the CL loss can be represented as following:
\begin{equation}
    \mathcal{L}_{\text{CL}}=-\frac{1}{2N}\sum_{i=1}^N\big(\log\frac{\tilde{s}_{2i-1,2i}}{\tilde{s}_{2i-1}} + \log \frac{\tilde{s}_{2i,2i-1}}{\tilde{s}_{2i}}\big).
\end{equation}

\subsection{Relative Position Learning}
Due to the considerations previously mentioned in Section~\ref{sec:1}, the second goal of our designed unsupervised learning task is to endow the base model with the capability to predict relative position (RP).
Specifically, we abstract RP prediction into a function: taking two neighboring images $\mathbf{M}_A$ and $\mathbf{M}_B$ as input, predicting the position of $\mathbf{M}_B$ relative to $\mathbf{M}_A$ as output.

CL can only yield the capability to answer whether two images are neighboring, but not the capability to answer what the RP between two images is.
In CL, two neighboring images $\mathbf{M}_A$ and $\mathbf{M}_B$ can be viewed as two variants generated by data augmentations (resize and crop) from a larger image containing $\mathbf{M}_A$ and $\mathbf{M}_B$ concurrently, and hence their representations are constrained to be of high cosine similarity by CL.
In other words, the value of $\textbf{sim}(\mathbf{M}_A,\mathbf{M}_B)$ is very close to $1$.
Therefore, with the model well trained by CL, we can easily estimate the probability that two images are neighboring (\textit{i.e.,} derived from the same larger image) according to the calculated cosine similarity between their representations.
However, the constraints induced by CL is completely relative-position-agnostic.
Specifically, the representations of $\mathbf{M}_A$ and $\mathbf{M}_B$ are equally constrained regardless of whether $\mathbf{M}_B$ is to the left or right of $\mathbf{M}_A$.
Therefore, the representations learned by CL do not carry any information about RP, ultimately resulting in the model failing to handle RP prediction.
As we aforementioned in Section~\ref{sec:1}, this may be inconsistent with the reality reported in recent studies~\cite{konen2008two,kravitz2010high,sereno2011population,zachariou2014ventral} that VVS also contributes to location perception.


To compensate for the above functional defect of CL, we specially design relative position learning (RPL), which makes the base model explicitly learn the knowledge about RP.
Specifically, for each sampled image $\mathbf{M}_i$ in current training round, we first divide it into four equal-sized $2\times 2$ blocks denoted by $\mathbf{X}_i^0$, $\mathbf{X}_i^1$, $\mathbf{X}_i^2$, and $\mathbf{X}_i^3$ respectively.
Then we randomly select two unequal numbers from $\{0,1,2,3\}$, denoted by $a_i$ and $b_i$ ($a_i\neq b_i$), and obtain the label $d_i$ indicating the RP of $\mathbf{X}_i^{b_i}$ to $\mathbf{X}_i^{a_i}$.
Note that the value of $d_i$ is an integer in the range of $0$ to $7$, indicating eight directional categories (left, right, upper, lower, upper-left, upper-right, lower-left, and lower-right) respectively.
An additional learnable classifier head $h:\mathbb{V}^{2n}\rightarrow\mathbb{V}^8$ is involved in RPL, which is an MLP with 8-dimensional softmax outputs.
The classifier takes the concatenation of two image features extracted by the base model $f$ as inputs, and outputs the predicted probabilities of the eight directional categories.
For convenience, we define $\tilde{d}_{i}=h(f(\mathbf{X}_i^{a_i})\oplus f(\mathbf{X}_i^{b_i}))$, where $\oplus$ represents vector concatenation.
Finally, our RPL loss can be represented as following:
\begin{equation}
    \mathcal{L}_{\text{RPP}}=\frac{1}{N}\sum_{i=1}^N l(\tilde{d}_{i},d_i),
\end{equation}
where $l$ is the cross entropy loss as following:
\begin{equation}
    l(\tilde{d}_{i},d_i)= \frac{1}{8}\sum_{j=0}^{7} \sigma(d_i=j)\cdot\log(\tilde{d}_{i,j}).
\end{equation}
Note that $\tilde{d}_{i,j}$ is the $j$-th element of $\tilde{d}_{i}$, and $\sigma$ is the indicator function. $\sigma(e)=1$ if $e$ is true; otherwise, $\sigma(e)=0$.

After RPL, the RP information is embedded into the features extracted by the base model $f$, further utilized by the classifier head $h$ to predict RP.
From this aspect, the base model $f$ has obtained the capability of RP prediction.

\subsection{Method Overview}
For training our base model $f$ to concurrently handle the two visual functions (object recognition and RP prediction), we craft an unsupervised dual-task learning.
Specifically, the overall loss of our proposed dual-task learning is the linear combination of the CL loss $\mathcal{L}_{\text{CL}}$ and the RPL loss $\mathcal{L}_{\text{RPL}}$, as following:
\begin{equation}
    \mathcal{L}=\mathcal{L}_{\text{CL}}+\alpha\cdot\mathcal{L}_{\text{RPL}},
\end{equation}
where $\alpha$ is a hyper-parameter to balance the relative weight between two parts.
Note that base model $f$, projection head $g$, and classifier head $h$ are optimized simultaneously by minimizing $\mathcal{L}$.

%% file: sections/Results.tex
\section{Experiments}
For thoroughly testing our proposed new unsupervised task-driven methods to model VVS, we conduct extensive experiments.
In this section, we first introduce some basic experimental settings, and then demonstrate the results with detailed analyses.
The source code of our experiments is available at \url{https://github.com/rdz98/Unsup-VVS}.

\subsection{Datasets}
Our experiments involve one image dataset (STL-10) and two neural datasets (Macaque-V1/V2 and Macaque-V4/IT).
In the following, we introduce the three datasets respectively.

\textbf{STL-10}\footnote{\url{https://cs.stanford.edu/~acoates/stl10/}.}. The STL-10 dataset~\cite{coates2011analysis} is a widely used public benchmark dataset for image classification, containing 10 object classes with 5,000 labeled training images, 8,000 labeled test images, and 100,000 unlabeled images. Each image is $96\times 96$ pixels in color.

\textbf{Macaque-V1/V2}\footnote{Both of the two neural datasets are publicly accessible on Brain-Score~\cite{SchrimpfKubilius2018BrainScore,Schrimpf2020integrative}.}. The neural responses of anesthetized macaque monkeys to 2,700 texture image stimuli were recorded in this dataset~\cite{freeman2013functional}, involving 102 V1 neurons and 103 V2 neurons. The texture image stimuli consists of 20 repetitions of 135 naturalistic or noise samples from different texture families.

\textbf{Macaque-V4/IT}\footnotemark[\value{footnote}]. The neural responses of alert rhesus macaque monkeys to 3,200 synthetic images were recorded in this dataset~\cite{majaj2015simple}, involving 88 neurons in V4 and 168 neurons in IT. The images were generated by adding a 2D projection varying position, size and pose concomitantly of a 3D object model to a random natural background. There were 64 different 3D object models (8 basic-level categories, each with 8 exemplars).

\begin{table*}
    \centering
    \caption{Task Accuracy and Brain Similarity with Different Balancing Weights}\label{tab:1}
    \begin{tabular}{l|cc|cccc}
        \toprule
        \multirow{2}*{Balancing Weight} & \multicolumn{2}{c|}{Task Accuracy (\%)} & \multicolumn{4}{c}{Brain Similarity} \\ \cline{2-7}
        ~ & IC & RPP & V1 & V2 & V4 & IT\\ \midrule
        $\alpha=0$ (PCL) & 82.60 & 14.19 & 0.2198 ± 0.0165 & 0.3367 ± 0.0143 & 0.5236 ± 0.0056 & 0.4771 ± 0.0029  \\ 
        $\alpha=0.00002$ & 85.61 & 35.55 & 0.2391 ± 0.0139 & 0.3330 ± 0.0192 & 0.5237 ± 0.0045 & 0.5027 ± 0.0025  \\ 
        $\alpha=0.0001$ & 85.68 & 80.44 & 0.2217 ± 0.0152 & 0.3439 ± 0.0141 & 0.5291 ± 0.0045 & 0.5015 ± 0.0033  \\ 
        $\alpha=0.0005$ & 85.74 & 86.05 & 0.2390 ± 0.0099 & 0.3456 ± 0.0204 & 0.5312 ± 0.0050 & \textbf{0.5045 ± 0.0035}  \\ 
        $\alpha=0.002$ & \textbf{86.34} & 91.43 & 0.2440 ± 0.0115 & 0.3412 ± 0.0177 & 0.5304 ± 0.0041 & 0.4998 ± 0.0032  \\ 
        $\alpha=0.01$ & 85.23 & \textbf{95.28} & \textbf{0.2500 ± 0.0113} & \textbf{0.3521 ± 0.0146} & \textbf{0.5413 ± 0.0045} & 0.4932 ± 0.0035  \\ 
        $\alpha=0.05$ & 81.91 & 93.25 & 0.2303 ± 0.0140 & 0.3510 ± 0.0134 & 0.5371 ± 0.0047 & 0.4817 ± 0.0027  \\ 
        \bottomrule
    \end{tabular}
\end{table*}

\subsection{Metrics}
After training the model by the unsupervised task on the image dataset, we adopt three metrics to respectively evaluate the model from three aspects: image classification accuracy, relative position prediction accuracy, and brain similarity.
Notably, the first two are computed on the image dataset and we refer to them collectively as task accuracy, while the last is computed on the neural datasets.

\textbf{Image Classification (IC) Accuracy.}
When evaluating IC accuracy, we drop projection head $g$ and classifier head $h$ while retain the base model $f$ only.
We froze the base model $f$, and train a new linear classifier by L-BFGS optimization algorithm on labeled training samples in the image dataset.
Specifically, for each labeled training sample $(\mathbf{M}_i,y_i)$, the new linear classifier takes $f(\mathbf{M}_i)$ as inputs, and takes $y_i$ as the ground-true category label.
After training the new linear classifier, we calculate its top-1 classification accuracy on labeled test set of the image dataset as our IC accuracy.

\textbf{Relative Position Prediction (RPP) Accuracy.}
When evaluating RPP accuracy, we drop projection head $g$ while retain the base model $f$ and classifier head $h$.
For each image $\mathbf{M}_i$ in the test set of the image dataset, we adopt the same approach used in RPL to generate a test sample consisting of $\mathbf{X}_i^{a_i}$, $\mathbf{X}_i^{a_i}$, and $d_i$.
We take $d_i$ as the ground-true label of RP, and take $h(f(\mathbf{X}_i^a)\oplus f(\mathbf{X}_i^b))$ as the predicted probabilities of the eight RP categories.
Without any further training, we directly calculate top-1 classification accuracy on the generated test samples as our RPP accuracy.

\textbf{Brain Similarity.}
When evaluating brain similarity, we drop projection head $g$ and classifier head $h$ while retain the base model $f$ only.
We calculate the layer-level brain similarity for each layer of the base model respectively.
Specifically, we first input all visual stimuli from the neural dataset into $f$ to obtain the layer outputs.
Afterwards, we fit the layer outputs to predict the corresponding neural responses recorded in the neural dataset by Partial Least Squares (PLS) regression.
Subsequently, for each neuron, we compute the Pearson correlation coefficient between predicted responses and recorded responses, and then correct it by the neuron noise ceiling.
Finally, we obtain the layer-level brain similarity as the median of the noise-corrected correlation coefficients.
The brain similarity of the whole base model is the maximum of the layer-level brain similarities across the layers in the base model.

\subsection{Experimental Settings}
\subsubsection{Parameters.}
Considering the significant success of deep residual convolutional neural networks~\cite{he2016deep} on various computer vision tasks, we adopt ResNet-18 as our base model $f$.
The extracted feature dimension $n=512$, the projection dimension $m=128$, and the batch size $N=512$.
Besides, both the projection head $g$ and the classifier head $h$ are 2-layer MLPs with hidden layer dimension of $512$.
In our unsupervised training on the image dataset, we adopt Stochastic Gradient Descent (SGD) optimizer with learning rate $1.5$, weight decay $1\times 10^{-6}$, and momentum $0.9$.
The unsupervised training lasts for $500$ epochs.

\subsubsection{Baselines.}
In~\cite{zhuang2021unsupervised}, contrastive learning yields the s-o-t-a brain similarity among existing unsupervised task-driven methods to model VVS.
In this paper, we call the method as pure contrastive learning (PCL), and employ it as a baseline for comparison with our approach. Actually, when setting the balancing weight $\alpha=0$, our approach degenerates into PCL.


\subsection{How does RPL affect task accuracy?}
To explore how does RPL affect task accuracy (including IC accuracy and RPP accuracy), we first perform our designed unsupervised training for seven times with different balancing weights ($\alpha=0$, $0.00002$, $0.0001$, $0.0005$, $0.002$, $0.01$, and $0.05$), and obtain seven models respectively.
Then we evaluate task accuracy of the seven models, and the results are shown in Table~\ref{tab:1}.
From the observation of task accuracy in the table, we can conclude the following three points:
\begin{enumerate}
    \item Increasing the weight of RPL (from $\alpha=0$ to $\alpha=0.002$) not only improves RPP accuracy significantly (from $14.19$ to $91.43$) but also improves IC accuracy slightly (from $82.60$ to $86.34$). This is consistent with recent finding~\cite{doersch2017multi,pinto2017learning} that combining multiple unsupervised tasks (CL and RPL) always leads to the improvement of the downstream performance (IC accuracy).
    \item When the weight of RPL is exceedingly small ($\alpha=0.00002$), the obtained model has poor RPP accuracy ($35.55\%$), but IC accuracy is still significantly enhanced compared to that with $\alpha=0$. This demonstrates that even though the model is not well-trained on RPL, RPL can help CL to learn better image representations, hence the better downstream performance.
    \item When the weight of RPL is too large ($\alpha=0.05$), the interference between two tasks arises, causing drops in both IC accuracy and RPP accuracy.
\end{enumerate}

\begin{table*}
    \centering
    \caption{Layer-level Brain Similarity to V1, V2, V4 and IT}\label{tab:2}
    \begin{tabular}{c|cc|c|cc}
        \toprule
        \multirow{2}*{Model Layer} & \multicolumn{2}{c|}{V1} & \multirow{2}*{Model Layer} & \multicolumn{2}{c}{V2}\\ \cline{2-3}\cline{5-6}
        ~ & PCL & Ours & ~ & PCL & Ours\\ \midrule
        layer1.0 & 0.1730 ± 0.0156 & \textbf{0.1987 ± 0.0094} & layer1.0 & 0.1945 ± 0.0197 & \textbf{0.2283 ± 0.0167}\\
        layer1.1 & 0.1676 ± 0.0152 & \textbf{0.2116 ± 0.0088} & layer1.1 & 0.2248 ± 0.0230 & \textbf{0.2636 ± 0.0159}\\
        layer2.0 & 0.2198 ± 0.0165 & \textbf{0.2445 ± 0.0128} & layer2.0 & 0.3130 ± 0.0162 & \textbf{0.3429 ± 0.019}5\\
        layer2.1 & 0.2131 ± 0.0179 & \textbf{0.2500 ± 0.0113} & layer2.1 & 0.2990 ± 0.0164 & \textbf{0.3270 ± 0.0187}\\
        layer3.0 & 0.1748 ± 0.0148 & \textbf{0.1904 ± 0.0117} & layer3.0 & 0.3249 ± 0.0148 & \textbf{0.3369 ± 0.0192}\\
        layer3.1 & 0.1633 ± 0.0143 & \textbf{0.1863 ± 0.0141} & layer3.1 & 0.3367 ± 0.0143 & \textbf{0.3521 ± 0.0146}\\
        layer4.0 & 0.1473 ± 0.0125 & \textbf{0.1674 ± 0.0116} & layer4.0 & 0.3028 ± 0.0131 & \textbf{0.3304 ± 0.0114}\\
        layer4.1 & \textbf{0.1538 ± 0.0088} & 0.1525 ± 0.0077 & layer4.1 & 0.3096 ± 0.0160 & \textbf{0.3185 ± 0.0136}\\
        \bottomrule
        \multirow{2}*{Model Layer} & \multicolumn{2}{c|}{V4} & \multirow{2}*{Model Layer} & \multicolumn{2}{c}{IT}\\ \cline{2-3}\cline{5-6}
        ~ & PCL & Ours & ~ & PCL & Ours\\ \midrule
        layer1.0 & 0.5058 ± 0.0046 & \textbf{0.5180 ± 0.0056} & layer1.0 & 0.3148 ± 0.0023 & \textbf{0.3224 ± 0.0034}\\
        layer1.1 & 0.5189 ± 0.0039 & \textbf{0.5338 ± 0.0046} & layer1.1 & 0.3440 ± 0.0021 & \textbf{0.3571 ± 0.0025}\\
        layer2.0 & 0.5236 ± 0.0056 & \textbf{0.5293 ± 0.0042} & layer2.0 & 0.4082 ± 0.0030 & \textbf{0.4158 ± 0.0041}\\
        layer2.1 & 0.5204 ± 0.0042 & \textbf{0.5413 ± 0.0045} & layer2.1 & 0.4356 ± 0.0037 & \textbf{0.4528 ± 0.0040}\\
        layer3.0 & 0.4509 ± 0.0055 & \textbf{0.4600 ± 0.0066} & layer3.0 & 0.4737 ± 0.0031 & \textbf{0.4861 ± 0.0036}\\
        layer3.1 & 0.4456 ± 0.0064 & \textbf{0.4586 ± 0.0062} & layer3.1 & 0.4771 ± 0.0029 & \textbf{0.4932 ± 0.0035}\\
        layer4.0 & \textbf{0.3404 ± 0.0067} & 0.3208 ± 0.0072 & layer4.0 & 0.4493 ± 0.0024 & \textbf{0.4593 ± 0.0039}\\
        layer4.1 & \textbf{0.3314 ± 0.0059} & 0.2746 ± 0.0076 & layer4.1 & \textbf{0.4408 ± 0.0035} & 0.4289 ± 0.0046\\
        \bottomrule
    \end{tabular}
\end{table*}

\begin{figure} 
\centering 
\includegraphics[width=0.9\linewidth]{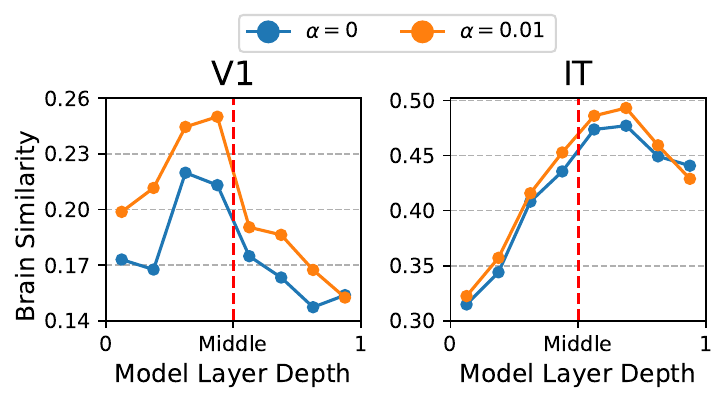} 
\caption{Layer-level Brain Similarity to V1 and IT} 
\label{fig:exp-f1} 
\end{figure}

\subsection{Can RP predictivity improve brain similarity?}
To investigate whether RP predictivity really improves the brain similarity, we further evaluate the brain similarity of the seven models with different balancing weights to four cortical regions (V1, V2, V4, and IT) of VVS.
The results are shown in Table~\ref{tab:1} as well.

It is evident that once RPL is included in our designed unsupervised task, the brain similarity to four visual cortical regions will be generally higher.
When setting $\alpha=0.01$, our method significantly outperforms PCL ($\alpha=0$), which enhances the brain similarity to V1 by $13.74\%$ from $0.2198$ to $0.2500$, enhances the brain similarity to V2 by $4.57\%$ from $0.3367$ to $0.3521$, enhances the brain similarity to V4 by $3.38\%$ from $0.5236$ to $0.5413$, and enhances the brain similarity to IT by $3.37\%$ from $0.4771$ to $0.4932$.
Through the comparison, we can conclude our proposed method sets the new state-of-the-art.

Despite the above significant improvement of brain similarity, we can not directly conclude RP predictivity can improve the brain similarity from it.
As we demonstrated previously, RPL also improves IC accuracy while enhancing RP predictivity.
The better IC accuracy indicates the stronger ability of object recognition of the model, which may also result in higher brain similarity.
Hence, the improvement of brain similarity may not be caused by the improvement of RPP accuracy, but be caused by the improvement of IC accuracy.
Fortunately, when $\alpha=0.05$, due to the relatively large $\alpha$, the interference between two tasks makes current IC accuracy ($81.91$) lower than baseline IC accuracy ($82.60$).
This case rules out IC accuracy improvement but current brain similarity is still comprehensively higher than the baseline to all of the four visual cortical regions.
Therefore, from this case, we can conclude that RP predictivity can really improve the brain similarity.

\subsection{Layer-level Brain Similarity Analysis}
To explore the influence of our proposed method on brain similarity in more detail, we select the final activation layers of each residual block in our base model (ResNet-18 totally contains $4*2$ residual blocks) and demonstrate their layer-level brain similarity to four visual cortical regions with PCL ($\alpha=0$) and our method ($\alpha=0.01$) respectively. The results are shown in Table~\ref{tab:2}.
Note that from ``layer1.0'' to ``layer4.1'', the layer depth increases progressively.
From the observation of the results we can conclude that, compared to PCL, our method can generally enhance the brain similarity of all base model layers to all visual cortical regions, with only very few exceptions occurring in the two deepest layers.

Besides, for better understanding the variation of layer-level brain similarity as the layer depth increases in our base model, we focus on the most primary visual area V1 and the highest visual area IT among the four visual areas, and we can find that: (i) the higher brain similarity to V1 is mostly achieved in the shallow layers of our base model (the layers before the middle), while (ii) the higher brain similarity to IT is mostly achieved in the deep layers of our base model (the layers after the middle). This means our base model roughly matches VVS structurally. The shallow layers correspond to the low-level visual areas, and the deep layers correspond to the high-level visual areas. Furthermore, it is evident in the figure that, compared to PCL ($\alpha=0$), our method ($\alpha=0.01$) preserves good model-layer-to-brain-area correspondence while generally enhancing the brain similarity across all layers.

%% file: sections/Discussion.tex
\section{Conclusion and Future Work}
Based on existing findings that ventral visual stream also contributes to location perception, in this paper we propose a new task-driven method to model ventral visual stream unsupervisedly, which integrates relative position learning with contrastive learning. The experimental results demonstrate that relative position predictivity can significantly improve the brain similarity, which corroborates the previous findings from a computational perspective.

Despite the outstanding brain similarity we achieved, there is still room for improvement.
In our future work, we will explore some more biologically plausible base models such as spiking neural networks instead of conventional ResNet.
